\documentclass[12pt,preprint]{aastex}
\usepackage{mathrsfs}

\newcommand\ka{\kappa}

\newcommand\<{\langle}
\renewcommand\>{\rangle}

\newcommand\beq{\begin{equation}}
\newcommand\eeq{\end{equation}}
\newcommand\bea{\begin{eqnarray}}
\newcommand\eea{\end{eqnarray}}
\newcommand\bal{\begin{align}}
\newcommand\eal{\end{align}}

\newcommand\be{\bold{e}}

\newcommand\bk{\bold{k}}

\newcommand\br{\bold{r}}

\renewcommand\bal{\mbox{\boldmath$\alpha$}}

\newcommand\tu{\tilde{U}}
\newcommand\bka{\mbox{\boldmath$\kappa$}}

\begin{document}

\title{The nature of intrinsic fluctuations in cosmic diffuse radiation}

\author{Richard Lieu\altaffilmark{1} \& Bizhu Jiang\altaffilmark{2}}

\altaffiltext{1}{Department of Physics, University of Alabama, Huntsville, AL
35899.}

\altaffiltext{2}{Physics Department and Center for Astrophysics,
Tsinghua University, Beijing 100084, China.}

\begin{abstract}

The spatial and temporal noise properties of diffuse radiation is
investigated in the context of the cosmic microwave background
(CMB), although generic formulae that enable application to any
other forms of incoherent light of a prescribed energy spectrum are
also provided.  It is shown that the variance of fluctuations in the
density and flux consists of two parts.  First is a term from the
spontaneous emission coefficient which is the contribution from a
random gas of classical particles representing the corpuscular
(photon) nature of light.  Second is a term from the stimulated
emission coefficient which leads to a `wave noise', or more
precisely the noise arising from the superposition of many plane
waves of arbitrary phase - the normal modes of the radiation. The
origin of this second term has never been elucidated before.  We
discussed one application. In the spatially homogeneous
post-inflationary epochs when the universe was reheated to GUT
temperatures, the thermal CMB density fluctuations on the horizon
scale is of order the WMAP measured value of 10$^{-5}$. Beyond
(larger than) the horizon, the power spectrum of perturbations could
moreover assume the observed form of $P(k) \sim k$ if thermal
diffusion to equilibrium does not merely involve classical
particles, but also non-localized wave components.  The outcome of
our study clearly demonstrates that such components do indeed exist.

\end{abstract}

\noindent

\noindent {\bf 1. Introduction: radiation noise and Poisson
statistics}

The fluctuation properties of diffuse radiation is a topic of
increasing importance, especially with the advent of cosmological
probes like \texttt{COBE} and \texttt{WMAP}, as they measure the
density perturbations of the early universe that seeded the
formation of structures (Bennett et al 2003, Hinshaw et al 2007,
Spergel et al 2007).  Yet, aside from the extrinsic physical
processes that generated and grew these seeds, the cosmic microwave
background (CMB), along with other forms of diffuse sky background,
has its own intrinsic noise.  It is helpful to acquire a better
understanding of this noise, so that at the very least one has a
firm handle of what to expect even in the limit of a perfect
instrument and a smooth sky (the latter from the `extrinsic'
standpoint).

The motivation of our present paper is to take a close look at the
intrinsic properties of radiation noise, with the black body
radiation (BBR) as our prime focus, although generalization will be
made to include other types of diffuse emissions.  In the field of
optics, this investigation was initiated and vigorously pursued by
several authors during the decade of 1950 -- 60 (Purcell 1956,
Hanbury Brown and Twiss 1957, Mandel 1958), and in recent times the
work was continued mostly by researchers of ultrafast optics
(Trebino 1997). Our purpose here is to further develop and
articulate the findings of these authors in the context of
astrophysical observations and interpretations.

We begin by revisiting BBR itself, by considering lengthscales and
timescales short enough that any effects due to the Hubble expansion
can be neglected.  The probability of having the integer $n$ as
occupation number of some normal mode $j$ is governed by
Bose-Einstein (BE) statistics, as \beq P_n =
\left(1-e^{-\beta\hbar\omega_j}\right) e^{-n\beta\hbar\omega_j},
\label{Pn}\eeq where $\beta = 1/(k_B T)$. The mean occupation number
and its variance are, as a result, \beq \bar n_j =
\frac{1}{e^{\beta\hbar\omega_j}-1},~{\rm and}~ (\delta n_j)^2 =
\frac{e^{\beta\hbar\omega_j}}{e^{\beta\hbar\omega_j}-1}. \label{nj}
\eeq Similar parameters for the energy contribution from this mode
are $\bar\epsilon_j = \bar n_j  \hbar\omega_j$ and $(\delta
\epsilon_j)^2 = (\delta n_j)^2 \hbar^2 \omega_j^2$. Since modes
within any black body cavity of volume $V$ are independent, the
variance of the total energy $\epsilon$ in this volume is simply
$(\delta\epsilon)^2 = \sum_j (\delta \epsilon_j)^2$.

Now, taking the continuum limit and applying the usual phase space
density formula for the number of electromagnetic wave modes per
interval $d\omega$ of angular frequency \beq \sum_j \rightarrow 2
\times \frac{V}{(2\pi)^3} \int d^3 {\bf k},~{\rm or}~\frac{dn_{{\rm
mode}}}{d\omega} = 2 \times \frac{\omega^2V}{2\pi^2 c^3},
\label{sumj}\eeq where the factor of two represents the polarization
degrees of freedom, one obtains $$ \bar\epsilon = \frac{8\pi V (k_B
T)^4}{(hc)^3} \int_0^\infty \frac{x^3 dx}{e^x -1} = \frac{8\pi^5 V
(k_B T)^4}{15(hc)^3}, $$  and \beq (\delta\epsilon)^2 = \frac{8\pi
V(k_B T)^5}{(hc)^3} \int_0^\infty \frac{x^4 e^x dx}{(e^x-1)^2} =
\frac{32\pi^5 k_B^5 T^5 V}{15 h^3 c^3}. \label{deltaeps}\eeq Hence
the contrast in $\epsilon$, or the energy density $u$, is \beq
\left(\frac{\delta\epsilon}{\epsilon}\right)^2 = \left(\frac{\delta
u}{u}\right)^2 = \frac{15 h^3 c^3}{2\pi^5 V k_B^3 T^3} =
\frac{16k_B}{3S}, \label{eden}\eeq in agreement with the derivation
of Peebles 1993 ($S=4\bar\epsilon/(3T)$ is the entropy of the
radiation).

An important point of physics has to be made about eq. (\ref{nj}).
The variance $(\delta n_j)^2$ in $n_j$ is greater than the mean
$\bar n_j$, i.e. BBR noise does {\it not} precisely follow Poisson
statistics. Yet the common misconception that in the complete
(quantum) theory radiation noise arises solely from photon counting
may still be defended. Thus, when integrated over all frequencies,
eqs. (\ref{deltaeps}) and (\ref{eden}) give the distinctly
Poissonian behavior of $(\delta\epsilon/\epsilon)^2 \sim 1/N$ where
$N = \int d\bar n$ is the total number of photons in the cavity of
volume $V$. Also, for the high frequency modes $x=\hbar\omega/(k_B
T) \gg 1$, $(\delta n)^2 \approx \bar n$, i.e. the discrete nature
of radiation as randomly moving photons prevails there.  However, by
writing the variance as two terms\footnote{The decomposition of eq.
(6) into $n(n+1)$ was historically what prompted Einstein to
conjecture the phenomenon of stimulated emission}, \beq (\delta n)^2
= \bar n (\bar n +1),~{\rm or}~(\delta n)^2 = \frac{e^x}{(e^x -1)^2}
= \frac{1}{e^x -1} + \frac{1}{(e^x -1)^2}. \label{varterms}\eeq  we
see that the Poisson noise term $\bar n$ dominates at $x \gg 1$,
whereas the $\bar n^2$ (rightmost) term is significant only at low
frequencies $x \ll$ 1. Focussing upon the latter, we may apply the
$x \ll$1 limit to eq. (\ref{deltaeps}) and obtain, for a frequency
interval $\delta x$ in this regime, $(\delta\epsilon)^2 \sim x^2
\delta x/h^3 \sim \nu^2 \delta\nu/(k_B T)^3$, so that the variance
is no longer dependent upon $h$. Clearly here, we are dealing with
noise of a `classical' nature.  But what {\it exactly} is it? The
question has never been answered rigorously.

\vspace{2mm}

\noindent {\bf 2. The precise origin of the non-Poisson component
and its importance}

One could take the questions raised above as a red herring, or at
best pedagogical: what prevents one from simply appealing to
radiation as an ensemble of photons, which in turn are bosons with
the manifestly non-Poisson distribution of eq. (\ref{Pn})?  In other
words, one already has an exact theory of BE statistics that renders
any `analysis' like the previous paragraph quite obsolete.  The
answer is three-fold. Firstly, the application of BE statistics to
systems of various sizes is non-trivial even for BBR.  Results like
eqs. (\ref{deltaeps}) and (\ref{eden}) actually describe an
incoherent ensemble of $\sim V/(\Delta\lambda)^3$ `pure BE' systems
with $\Delta\lambda$ denoting the bandwidth of the spectrum (i.e.
the range of wavelengths within which most of the emitted flux
resides); each of such systems corresponds to an elementary cell in
the standard phase space quantization scheme. If e.g. the volume of
space in question is $\lesssim \lambda^3$, coherence effects will
play a role, modifying the variance from the form eqs.
(\ref{deltaeps}) or (\ref{eden}), see Mandel 1958 (also Purcell
1956; Hanbury Brown, and Twiss 1957). The same `partitioning'
phenomenon applies as well to intensity fluctuations $\delta I$ over
time intervals large or small compared with $1/\Delta\omega$.  Such
behavior deserves further investigation, which is one purpose of
this paper.

Secondly, from the standpoint of modeling cosmological structures
there is especially a need to press for better understanding of the
two terms in eq. (\ref{varterms}).  The primordial density
perturbations that seeded structure formation have hitherto been
characterized by the two-point correlation function $\xi (\br_1 -
\br_2)$, the Fourier transform of which is the power spectrum
$P(k)$, which describes how particles are positioned in space via a
locally random (Poisson) process. If, however, during the radiation
era the fluctuations have a component that cannot be described in
terms of the distribution of classical Poisson particles at all,
then the properties of $\xi (\br_1 - \br_2)$ must be inferred
without this perspective, i.e. an altogether different microscopic
treatment will be necessary instead.

Thirdly, in astrophysics and experimental optics one could easily
envisage other types of radiation, such as the diffuse X-ray
background, or simply {\it any} arriving signals of incoherent light
- often known as white light.  If the mean energy spectrum (more
precisely spectral energy density) $\bar {\cal U}(\omega)$ of the
radiation is known, can we infer the fluctuations in the energy
density at any length scale, or the intensity measured over any time
scale? Obviously, a clear picture of the origin of {\it all} the
noises is prerequisite to the solution.

To elaborate upon this third point, let the mean occupation number
per mode be $\bar n(\omega)$, so the mean energy density of all
modes is \beq \bar{\cal U} = \frac{1}{\pi^2 c^3} \int_0^\infty
\omega^2 \hbar\omega \bar n(\omega) d\omega = \int_0^\infty \bar
{\cal U}(\omega) d\omega, \label{meden}\eeq where, from eq.
(\ref{sumj}), \beq \bar {\cal U}(\omega) = \frac{\omega^3}{\pi^2
c^3} \bar n(\omega). \label{Uw}\eeq Now the distribution $n(\omega)$
assumes, for a given value $\bar n(\omega)$, the most likely
occurring form when its Entropy is maximized subject to this
constraint on the mean, irrespective of whether each mode is in
`internal equilibrium' or not (if it is, the `most likely' qualifier
will be unnecessary, see Jaynes 1957 for the complete theory). Under
the scenario of states being occupied with complete {\it a priori}
freedom, i.e. boson systems with no prior restriction on $n$, the
resulting $n(\omega)$ would have the variance of $[\delta
n(\omega)]^2 = \bar n (\omega) [\bar n(\omega) +1]$. Hence, as in
eq. (\ref{meden}), \beq (\delta {\cal U})^2 = \frac{1}{\pi^2 c^3 V}
\int_0^\infty \omega^2 (\hbar\omega)^2 {\bar n (\omega) [\bar n
(\omega) + 1)]} d\omega = \frac{1}{V} \int_0^\infty
\left[\hbar\omega {\cal U} (\omega) + \frac{\pi^2 c^3 {\cal U}^2
(\omega)}{\omega^2} \right] d\omega. \label{deltaU}\eeq If the
radiation is BBR, i.e. \beq {\cal U} (\omega) = U(\omega) =
\frac{\hbar\omega^3}{\pi^2 c^3} \frac{1}{e^{\beta\hbar\omega}
-1},\label{BBRU} \eeq it will be trivial to show that eq.
(\ref{deltaU}) is then the same as eq. (\ref{deltaeps}), with the
Poisson and non-Poisson (or `wave') noise given by the ${\cal U}
(\omega)$ integral and the ${\cal U}^2 (\omega)$ integral
respectively.

In spite of the elegance of the above derivation, one still ought to
seek a more substantive physical justification of eq. (\ref{deltaU})
that does not appeal to any line of reasoning based upon the `game
of chances' of Jaynes 1957.  We only resorted to such an argument
because of our lack of fundamental understanding of the non-Poisson
noise.

\vspace{2mm}

\noindent {\bf 3. A heuristic treatment of wave superposition noise}

The `$n^2$-noise' contribution to the total variance of BBR
fluctuations may further  be elucidated, even if still
heuristically, by Maxwell's wave model of the radiation.  Let BBR
modes exist in the usual manner within some volume $V$. The
normalized electric field of mode $j$ is written as \beq {\bf E}_j
({\bf r}, t) = E_j {\bf e}_j \exp i({\bf k}_j \cdot {\bf r} -
\omega_j t); {\bf E}_j ({\bf r}, t=0) = E_j {\bf e}_j \exp (i{\bf
k}_j \cdot {\bf r}), \label{Ej}\eeq where ${\bf e}_j$ is the
polarization unit vector. The total energy is $ \epsilon = uV = V
\sum_j |E_j|^2$. It seems reasonable to assume that $E_j$ are
uncorrelated random variables, with a gaussian distribution of zero
mean and variance $\<|E_j|^2\> \sim \<\epsilon_j\>/V$, where
$\<\epsilon_j\>$ is as defined in the text following eq. (\ref{nj}).
Thus, the variance $(\delta\epsilon)^2$ is the sum of the individual
$(\delta\epsilon_j)^2$, i.e. \beq (\delta\epsilon)^2 = V^2 \sum_j
[\<|E_j|^4\> - (\<|E_j|^2\>)^2].\label{var1} \eeq  For a complex
gaussian random variable of zero mean, the second last term is twice
the last term, so that \beq (\delta\epsilon)^2 = \sum_j
\<\epsilon_j\>^2 = \frac{8\pi V(kT)^5}{(hc)^3} \int_0^\infty
\frac{x^4 dx}{(e^x-1)^2},\label{var2} \eeq which is the integrated
contribution to the total variance from the `non Poisson'
(rightmost) term of eq. (\ref{varterms}).

It should be clarified that classically the random complex nature of
the field amplitude $E_j$ is due to the wave phase $\phi_j$ (which
changes from mode to mode without correlation) rather than the
photon counting noise, which is why the variance is not associated
with Poisson statistics. In fact, we can rewrite eq. (\ref{meden})
as \beq {\bf E}_j ({\bf r}, t) = |E_j| {\bf e}_j \exp i({\bf k}_j
\cdot {\bf r} - \omega_j t + \phi_j), \label{Ejrt}\eeq where
$|E_j|^2 = U(\omega_j) $ and $U$ is as defined in eq. (\ref{BBRU}).
In the next section we proceed to perform a more rigorous
calculation of this wave noise (or, better still, phase noise), by
first attempting just the one dimensional problem of intensity
fluctuations at various time scales.

\newpage

\noindent {\bf 4. Intensity fluctuation due to random mode phases,
the coherence length of BBR}

The total intensity (or flux, in ergs~cm$^{-2}$~s$^{-1}$) of, for
simplicity, a collimated beam of BBR arriving at the observer's
point ${\bf r} = 0$ is given by \beq I(\tau) = \frac{1}{\tau}
\int_{-\tau/2}^{\tau/2}~c|{\bf E}(t)|^2 dt,\label{Itau}\eeq where
$\tau$ is the time interval over which the intensity is measured,
and \beq {\bf E}(t) = \sum_j {\bf E}_j(t),\label{Et}\eeq
with\footnote{We shall see in section 6 that the validity of
choosing a fixed point ${\bf r} =$ 0 to do the calculation requires
the detector to be centered at this point, and of size $\lesssim$
one radiation wavelength (to maintain spatial coherence) which in
the case of 3K light is $\sim$ 3 cm.} ${\bf E}_j(t) = {\bf E}_j
({\bf r}=0,t)$ and the right side is as defined in eq. (\ref{Ejrt}).

We shall soon see that $I(\tau)$ is a sum of two terms.  One is a
constant `DC' term, independent of $\tau$ in particular, while the
other is a random `AC' term of zero mean, the variance of which
decreases with increasing $\tau$.  This second term causes $I(\tau)$
as evaluated by means of eq. (\ref{Itau}) with different `time
origins' to fluctuate from one interval of duration $\tau$ to the
next.  The behavior is actually well known in optics (Trebino et al
1997), and is a consequence of the Uncertainty Principle.   To see
this readily, one could return to eq. (\ref{Ejrt}) and set \beq
\phi_j = 0~~{\rm for~all}~j. \label{phij}\eeq The sum over $j$
performed in eq. (\ref{Et}) to obtain the total wave electric field
would then be, in essence, a Fourier inverse transform of the Planck
function (more precisely $\sqrt{U(\omega)}$).  The result is a
`pulse' of duration $\Delta t \sim 1/\Delta\omega$, where
$\Delta\omega$ is the spectral width of $U(\omega)$, as is apparent
when one plots the intensity profile $I(\tau)$ against $\tau$ with
high time resolution, Figure~1. This scenario is known in the field
of Fourier optics as a `transform limited pulse', viz. the
Uncertainty Principle is satisfied in the `minimal wave packet'
sense of $\Delta\omega \Delta t \sim 1$, and it is attained by
setting all the mode phases to the same value - the highest degree
of spectral phase coherence that can only be found in laser light.

If, as is normally the case of ordinary `white light', the normal
modes have more complicated phase relationships than eq.
(\ref{phij}), the corresponding pulse will in general last longer
than the minimum duration, i.e. it will satisfy $\Delta\omega \Delta
t > 1$.  When the other end of the extreme is reached, such that the
phases $\phi_j$ exhibit the maximum complexity by assuming
uncorrelated random numbers as their values, one is dealing with
`ideal white light' (BBR falls under this category), viz. an
arbitrarily long and continuous pulse\footnote{This is illustrated
in Figure 11 of
\texttt{http://en.wikipedia.org/wiki/Coherence\_(physics).}} with
$\Delta\omega \Delta t \gg 1$ (Trebino et al 1997).  Nevertheless,
this continuous light train is in detail broken into many `spikes'
of all widths, ranging from one period of electric field oscillation
at the BBR peak frequency to the full span of $L/c$.  Such `spikes'
are to be realized in the `AC' term of the previous paragraph.

To substantiate these points, we numerically computed eqs.
(\ref{Ejrt}) through (\ref{Et}) at two different resolutions $\tau$,
one beneath the reciprocal bandwidth of the 3K BBR spectrum (i.e.
$\tau \ll 1/\Delta\omega$) and the other well above it, this time
assigning random phases $\phi_j$  to each mode $j$ as is appropriate
for the BBR.  The light intensity averaged over temporally
contiguous intervals\footnote{To obtain $I(\tau)$ for many of these
intervals of $\tau$ we need to repeatedly compute eq. (\ref{Itau})
using a fixed set of random phases but shifting the time of the
mid-point of the integration limits by the amount $\tau$ for every
interval, alternatively one can keep re-randomizing the phases on a
per interval basis.} of $\tau$, $I(\tau)$, is indeed now seen in
Figure 1c and Figure 2 to be continuous, yet it is equally obvious
that there are hardly any changes in $I$ on the scale of $\tau
\lesssim$ 10$^{-13}$ s; when $\tau$ is larger, at e.g.  $\tau =$
10$^{-10}$ s, noises then become apparent. In fact, as will be
proved below, variations occur on all scales inside the range
\beq\frac{1}{\Delta\omega} \lesssim \tau \lesssim
\frac{1}{\delta\omega} \label{taurange}\eeq and with properties
determined solely by the BBR spectrum $B(\omega)$ and independent of
the mode quantization frequency $\delta\omega$ used in our
simulation. For values of $\tau$ approaching the left inequality,
the light curve becomes relaxed and regular, eventually altogether
steady when $\tau \lesssim 1/\Delta\omega$ as Figure 2a reveals.
This is consistent (again) with the Uncertainty Principle, viz. on
timescales $\Delta t < 1/\Delta\omega$ there can be no randomness in
the signal.  Far from this limit, the intensity turns erratic as the
noise decorrelates from one resolution element $\tau$ to the
next\footnote{Figure 2b bears resemblance to the fluctuation pattern
of `white light' from incoherent mode superposition as depicted in
\texttt{http://en.wikipedia.org/wiki/Image:Spectral\_coherence\_continuous.png.}},
Figure 2b.  It should also be mentioned that the right inequality of
eq. (\ref{taurange}), viz. $\delta\omega \lesssim 1/\tau$, is
equivalent to the constraint imposed by the periodic boundary
condition as applied to a cube of side $L=c\tau$.

\begin{figure}[!h]
\begin{center}
\includegraphics[angle=0,width=6.5in]{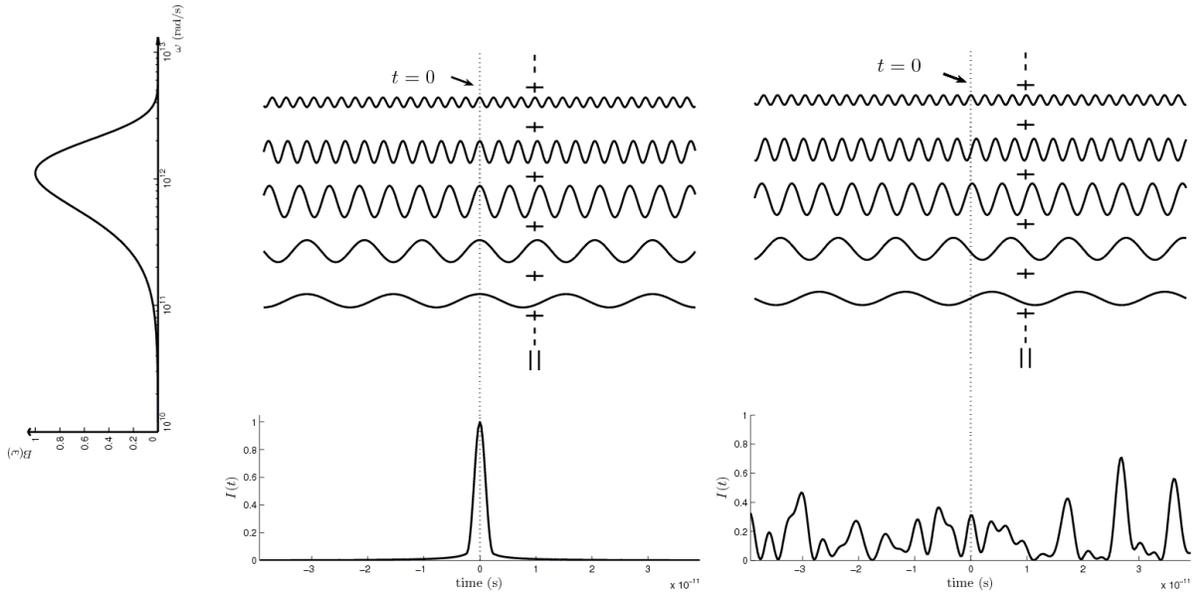}
\vspace{0.0mm}
\end{center}
\caption{Two simulated intensity profiles (arbitrary units and
photon Poisson noise excluded) of unpolarized light with the 3K
black body spectrum of 1a (extreme left). The detector is assumed to
be $\lesssim$ 1 cm, or one radiation wavelength, in width and
length, to avoid complications due to spatial incoherence.
Constituent normal modes are spaced $\delta\omega =$ 10$^{10}$ rad/s
apart.  1b (middle): all the modes are in phase, resulting in a
single pulse which is the Fourier transform of the spectrum. The
spectral width $\Delta\omega$ of 1a and the pulse width $\Delta t$
of 1b satisfy $\Delta\omega \Delta t \sim 1$. This is the scenario
of a `transform limited pulse' (or `minimal wave packet'), and
occurs only when the modes have full phase correlation.  1c (right):
the modes are randomly phased without any correlation among them;
here we see an enduring light curve with aperiodic spikes and
structures on all scales. Both profiles were constructed using eqs.
(\ref{Ejrt}) through (\ref{Et}), with a timing resolution of $\tau
=$ 10$^{-14}$ s.} \vspace{0.0cm}
\end{figure}

\begin{figure}[!h]
\begin{center}
\includegraphics[angle=0,width=3.2in]{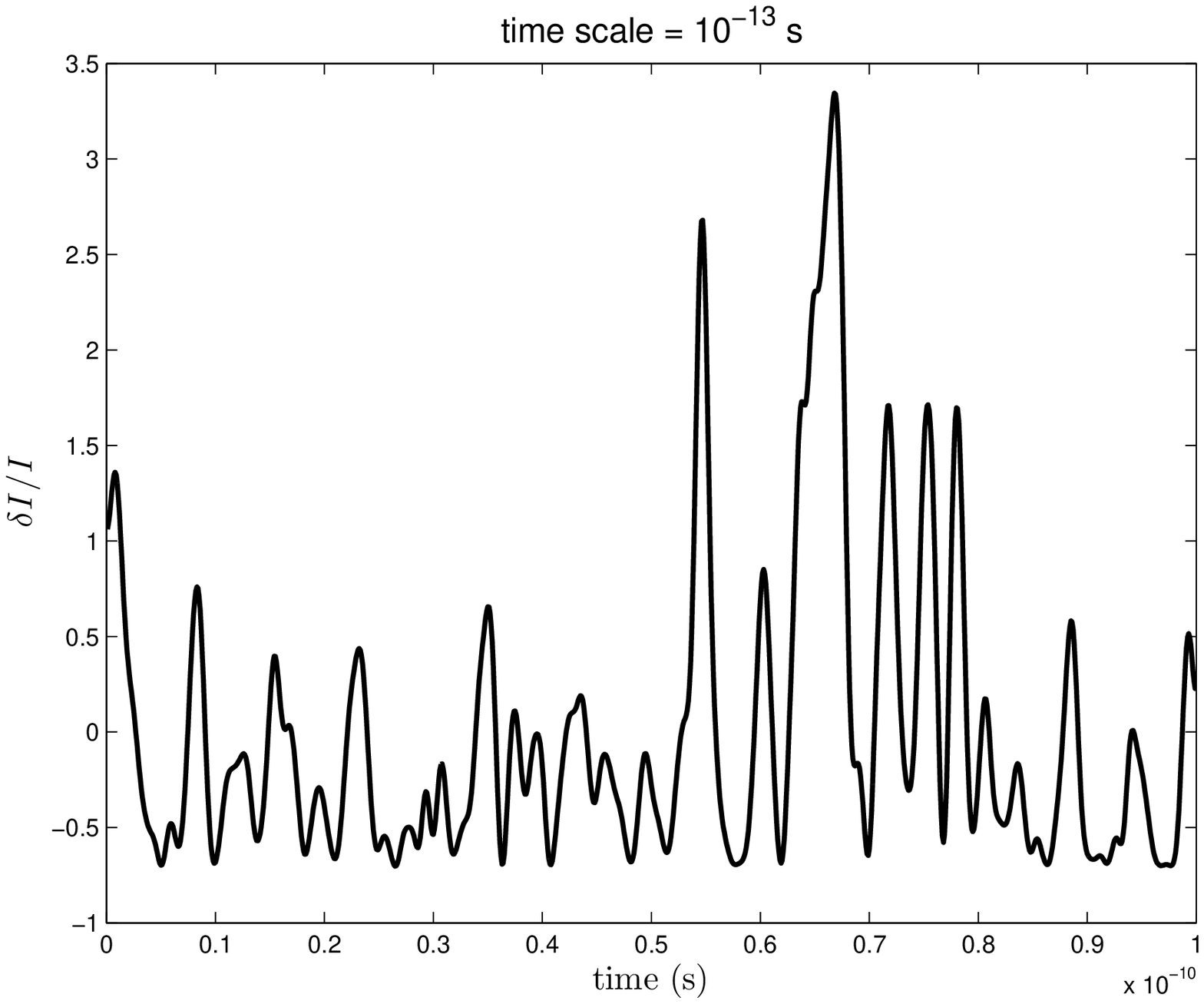}
\includegraphics[angle=0,width=3.2in]{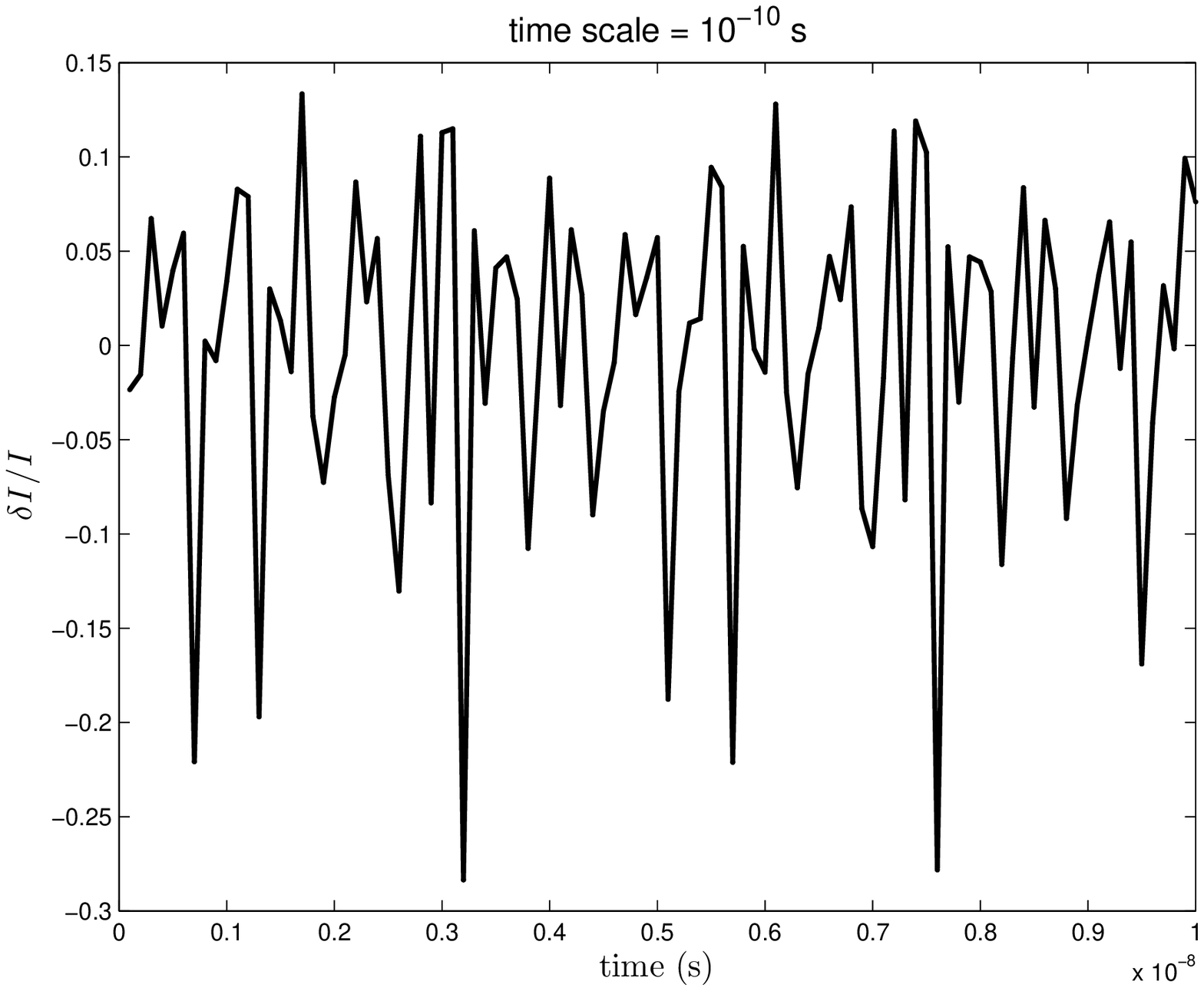}
\vspace{0.0mm}
\end{center}
\caption{Intensity fluctuations (phase noise component) of 3K BBR as
seen in two resolutions, 10$^{-13}$ s and 10$^{-10}$ s. Mode
frequency spacing adopted by our simulation is $\delta\omega =$
10$^{10}$ rad/s as in Figure 1.} \vspace{0.0cm}
\end{figure}

\begin{figure}[!h]
\begin{center}
\includegraphics[angle=0,width=5in]{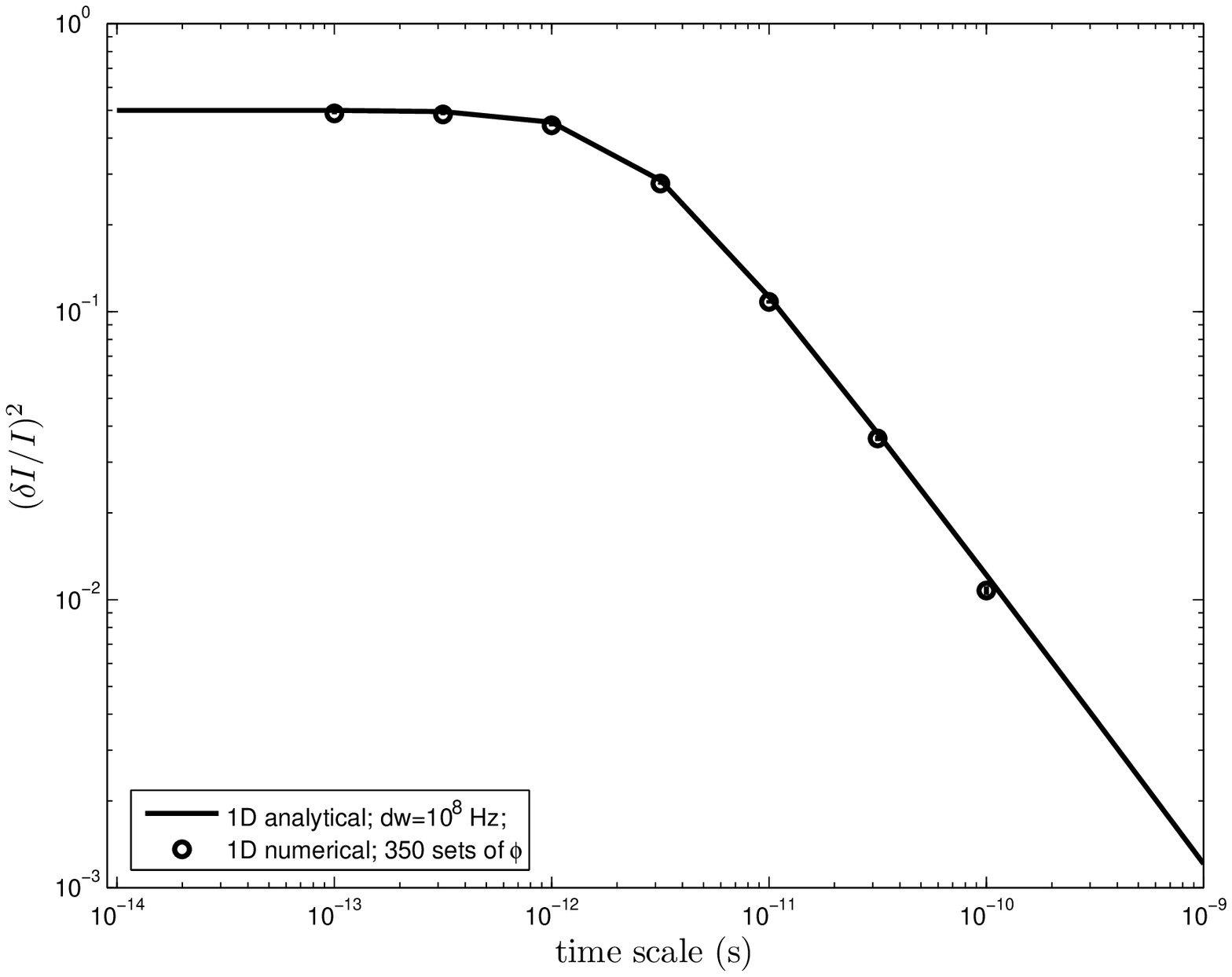}
\vspace{0.0mm}
\end{center}
\caption{Power spectrum of temporal intensity fluctuations of 3K
BBR, excluding photon Poisson noise.  The uncertainties in the
numerical simulations are $\lesssim$ the size of the circles, and
are determined by generating data from 350 different sets of random
phases $\phi_j$, then evaluating the scatter among the variances
that result from each. The mode quantization adopted throughout is
$\delta\omega =$ 10$^{10}$ rad/s.} \vspace{0.0cm}
\end{figure}

Our simulations for Figures 1 and 2 were pursued with a mode
quantization $\delta\omega =$ 10$^{10}$ rad/s.  Further beyond
plotting the intensity at the two $\tau$ values discussed, we
calculated from such plots the normalized variance $(\delta I/I)^2$
(where $I$ is the `global' mean intensity which  is a constant
independent of $\tau$) at each scale between $\tau =$ 10$^{-14}$ s
and $\tau =$ 10$^{-10}$ s. The results are shown in Figure 3 as the
data points of a power spectrum (in the WMAP sense of the words,
i.e. fluctuation power at various scales). Towards $\tau =$
10$^{-10}$~s the variance $\sim 1/\tau$, suggesting (as already
said) that neighboring bins of $\tau$ decorrelate when $\tau$
becomes large.  The argument is as follows.  From eq. (\ref{Itau})
the average intensity over some long interval ${\cal T}= N\tau$ may
be written as  \beq I_{{\cal T}} = c \times \frac{1}{{\cal T}}
\left[\int_0^\tau |{\bf E}(t)|^2 dt + \int_\tau^{2\tau} |{\bf
E}(t)|^2 dt + \cdots + \int_{(N-1)\tau}^{N\tau} |{\bf E}(t)|^2
dt\right]. \label{IT}\eeq If each of the N terms on the right side
fluctuates freely with a variance $\sigma_\tau^2$, we will have
$(\delta I_{{\cal T}})^2 = N\sigma_\tau^2/{\cal T}^2 =
\sigma_\tau^2/(N\tau^2)$; but since $(\delta I_\tau)^2 =
\sigma_\tau^2/\tau^2$ we obtain $(\delta I_{{\cal T}})^2 = (\delta
I_\tau)^2/N$, or ${\cal T} (\delta I_{{\cal T}})^2 = \tau (\delta
I_\tau)^2 =$ constant.  Hence $(\delta I_{{\cal T}})^2 \sim 1/{\cal
T}$.  This trend holds as long as we compare two timescales in which
the noise in the larger bin is simply the incoherent reinforcement
of those in the smaller ones.

At $\tau \lesssim$ 10$^{-11}$ s, however, we see in Figure 3 a
transition towards different behavior. As Figures 1c and 2a showed,
the intensity changes over short intervals are regular and highly
correlated, and variances no longer add. Eventually, $(\delta
I_{{\cal T}})^2 = (\delta I_\tau)^2$ and the power spectrum flattens
at the maximum, i.e.  no `extra' fluctuations may be found by
measuring the intensity using more and more accurate clocks, so
$\delta I$ does not rise any further.  From Figure 3, one finds that
the cross over point from constant $\delta I_\tau$ to the $\delta
I_\tau \sim 1/\sqrt{\tau}$ dependence is at $\tau = \tau_c\approx$
10$^{-10}$ s. This length may be used to define the coherence scale
of BBR at $T =$ 3~K.  Beneath $\tau_c$ (or $c\tau_c =$ 3 cm) a
wavetrain of BBR exhibits good phase correlation, above it the phase
becomes randomized.  In another manner of expression, 3K radiation
is capable of maintaining the integrity of its electromagnetic
oscillation pattern for $\omega_{{\rm max}} \tau_c \approx$ 100
cycles.  The overall picture is that one could divide space-time
into many cells of this critical dimension, and assume coherence
within each cell.  Thus e.g. the point raised earlier in footnote 2
may now be understood in better terms: the power spectrum simulated
and derived in this section is valid only for a detector size of
$\lesssim$ 3~cm because we have not taken into account the
additional effects of spatial incoherence.

Finally, it is also clear that the salient features of this section
applies to diffuse radiation of {\it any} spectrum $B(\omega)$.

\vspace{2mm}

\noindent {\bf 5. Analytical proof of temporal phase noise variance}

The outcome of section 4 is not only relevant to BBR, but to
incoherent light (or `white light') with any broadband spectrum.  A
more rigorous mathematical proof to support the simulations is
clearly desirable.

The intensity $I(\tau)$ at resolution $\tau$, eq. (\ref{Itau}), has
a `DC' and an `AC' contribution as mentioned after eq. (\ref{Et}).
These may now be written as \beq I_{{\rm DC}} = \sum_j B_j;~~I_{{\rm
AC}} (\tau) = \sum_{{\bf k}, m} \sum_{(\bk',n)\neq (\bk,m)}
\be_{\bk,m} \cdot \be_{\bk',n} \sqrt{B_{k} B_{k'}}
\frac{\sin\left[\frac{(\omega_k - \omega_{k'})
\tau}{2}\right]}{(\omega_k - \omega_{k'}) \tau}
e^{i\phi_{\bk,\bk',m,n}}, \label{IDCAC}\eeq where
$\phi_{\bk,\bk',m,n} = \phi_{\bk,m} - \phi_{\bk',n}$ ($m,n=$ 1,2 are
polarization labels), $B_k = cU_k = \hbar k c^2/(e^{\beta\hbar ck} -
1)$. Evidently $I_{{\rm DC}}$ is just a constant. Pairing each term
in the series of $I_{{\rm AC}}$ with its complex conjugate, the
$e^{i\phi}$ factor in eq. (\ref{IDCAC}) is then replaced by
$2\cos\phi$, and the $\sum_{\bk'}$ procedure is to be done with the
understanding that only one member of every $(\bk,\bk')$ and
$(\bk',\bk)$ pair should appear in the sum. The mean of $I_{{\rm
AC}}$ is evidently zero. The variance is, in the continuum
representation where $B(\omega)= cU(\omega)$ with the latter given
by eq. (\ref{BBRU}), \beq (\delta I_\tau)^2 = \frac{2}{\tau^2}
\int_0^\infty d\omega B(\omega) \int_0^\infty d\omega' B(\omega')
\frac{\sin^2\left[\frac{(\omega' - \omega) \tau}{2}\right]}{(\omega'
- \omega)^2}, \label{deltaIT}\eeq after applying a correction factor
of 1/2 to free the $\sum_{\bk'}$ step from the awkward `permutation'
restriction above; also the relation $\langle\cos^2\phi\rangle =$
1/2 was used
for the random phase differences $\phi$, and the sum over random
polarizations follows the standard formula \beq \sum_m \sum_n ({\bf
e}_{{\bf k},m} \cdot {\bf e}_{{\bf k'},n})^2 = 1 + \left(\frac{{\bf
k} \cdot {\bf k'}}{kk'}\right)^2 = 2 \label{summn}\eeq ($\bk \cdot
\bk' \approx kk'$ because $\omega' \approx \omega$ as we saw, and
$\bk \| \bk'$ for a collimated beam), this sum replaces the factor
of four from the helicity enumeration in the two phase space factors
that accompany $B_k$ and $B_{k'}$ as they become $B(\omega)$ and
$B(\omega')$.

A plot of $(\delta I_\tau/I)^2$ against $\tau$, using eq.
(\ref{deltaIT}) and the 3K Planck function of eq. (\ref{BBRU}), is
given by the solid line of Figure~3, where excellent agreement with
data from the numerical simulation of section 3 is apparent. There
are two obvious limiting cases in which eq. (\ref{deltaIT})
simplifies. First is when $\tau \ll 1/\Delta\omega$, the reciprocal
of the BBR spectral width. In the second integral we may now write
$\sin(\omega'-\omega)\tau \approx (\omega'-\omega)\tau$, and
$(\delta I_\tau)^2$ becomes a constant, as can be seen from the
flattened top portion of the variance curve in Figure~3.  Second is
when $\tau \gg 1/\Delta\omega$.  Here it is tempting to use the
property $\sin^2\psi \leq 1$ to recast eq. (\ref{deltaIT}) as \beq
(\delta I_\tau)^2 < \frac{2}{\tau^2} \int_0^\infty d\omega B(\omega)
\int_0^\infty d\omega' \frac{B(\omega')}{(\omega' - \omega)^2}
\label{deltaITmax}\eeq to conclude that $(\delta I_\tau)^2$ must
asymptotically fall with $\tau$ at least as steeply as $1/\tau^2$.
This is wrong for the following reason.  The function
$B(\omega')/(\omega' - \omega)^2$ always diverges at some point in
the $\omega'$ integration in such a way as to yield an infinite
upper limit to $(\delta I_\tau)^2$, of the form $\int_0^\infty
d\zeta/\zeta^2$.

Instead, to arrive at the correct asymptote one must observe that
for $\tau \gg 1/\Delta\omega$ the $\int d\omega'$ step in eq.
(\ref{deltaIT}) converges rapidly: within some interval $\omega_0 =
|\omega' - \omega| \ll \Delta\omega$, enabling us to write
$B(\omega') \approx B(\omega)$ and take the function $B(\omega)$ out
of the $\omega'$ integration, i.e. eq. (\ref{deltaIT}) reads \beq
(\delta I_\tau)^2 = 2\int_0^\infty d\omega [B(\omega)]^2
\int_{-\omega_0}^{\omega_0} d\tilde\omega
\frac{\sin^2\left(\frac{\tilde\omega
\tau}{2}\right)}{\tilde\omega^2\tau^2} = \frac{2}{\tau}\int_0^\infty
d\omega [B(\omega)]^2 \int_0^\infty \frac{\sin^2 \xi}{\xi^2} d\xi
\propto \frac{1}{\tau},  \label{deltaIT2}\eeq for $\tau \gg
1/\Delta\omega$, where $\tilde\omega=\omega'-\omega$ and
$\xi=\tilde\omega \tau/2$. Note that the integral over $\xi$ was
taken to $\xi = \infty$, resulting in the pure dimensionless number
of $\pi/2$, because when $\tau$ is large convergence to this number
only requires an excursion of $\tilde\omega$ across a very small
passband ($\sim\omega_0$ in width) within the Planck spectrum.  The
asymptotic $1/\tau$ dependence of the variance is, finally, $$
\left(\frac{\delta I_\tau}{I}\right)^2 = \frac{225h}{2\pi^8 \tau kT}
\int_0^\infty \frac{x^6 dx}{(e^x -1)^2} = 7.3 \times 10^2 \times
\frac{h}{\pi^8 \tau kT} = 0.0123 \left(\frac{\tau}{10^{-10}~{\rm
s}}\right)^{-1}~{\rm for}~T=3~{\rm K}, $$ in agreement with the
simulated data of Figure 3, and consistent with the rationale
presented around eq. (\ref{IT}) of large scale fluctuations being
fed by incoherent noises at smaller scales (see also Mandel 1958).

It is worth a mention in passing that one can also pursue the
simpler problem of the variance in $E(\tau)$, where \beq
E(\tau)=\frac{1}{\tau} \int_{-\tau/2}^{\tau/2}~{\rm Re} [E(t)] dt =
\sum_j 2\sqrt{B_j} \frac{\sin\left(\frac{\omega_j
\tau}{2}\right)}{\omega_j \tau} \cos\phi_j \label{Etau}\eeq is the
magnitude of the BBR electric field averaged over the time $\tau$ (a
quantity that obviously fluctuates about zero mean), viz. \beq
(\delta E_\tau)^2 = \sum_j 2 B_j \frac{\sin^2\left(\frac{\omega_j
\tau}{2}\right)}{\omega_j^2 \tau^2} , \label{deltaEtau}\eeq with the
last result obtained by applying $\langle\cos^2\phi\rangle =$ 1/2 as
before. The main difference from the $(\delta I_\tau)^2$ calculation
of eqs. (\ref{deltaIT}) through (\ref{deltaITmax}) is that this
time, we can actually follow the approach of eq. (\ref{deltaITmax})
without making a mistake, viz. by writing eq. (\ref{deltaEtau}) as
\beq (\delta E_\tau)^2 < \frac{2}{\tau^2} \int_0^\infty d\omega
\frac{B(\omega)}{\omega^2}. \label{deltaEtaumax} \eeq  The key point
is that there is no longer any infra-red divergence in the
integrand: at low frequencies $B(\omega)/\omega^2$ remains finite
due to the Rayleigh-Jeans Law $B(\omega) \sim \omega^2$ for small
$\omega$. The large scale r.m.s. electric field, then, can be
rationalized in terms of the arguments presented near eq.
(\ref{deltaITmax}), i.e.  it falls off at the rate of $1/\tau^2$ or
steeper, and the velocity of any charged particle $v = (eE_{{\rm
rms}}/m) \tau$ does not increase with $\tau$.  This is a reflection
of the thermal equilibrium environment of BBR: the long wavelength
fields have already been dissipated by re-absorption and ensuing
particle acceleration.

\vspace{2mm}

\noindent {\bf 6. Spatial contrast of the energy density of BBR -
the phase noise contribution}

We are now in a position to tackle the problem of spatial
fluctuations - the task identified in sections 1 and 2 of finding
the precise origin of the phase noise contribution to the spatial
variation of BBR energy density.  This time, we set in eq.
(\ref{Ejrt}) $t=$ 0, i.e. we take a snapshot of the radiation filled
space (more precisely defined as an interval of duration $\lesssim
1/\Delta\omega$ centered at the time origin) and examine the density
distribution everywhere.  The energy density $U(V)$ measured over
some cubical volume of side $L$ is given by \beq u(V) = \frac{1}{V}
\int |{\bf E}({\bf r})|^2 d^3 {\bf r}, \label{uR}\eeq where $V=L^3$.
Following the notation of sections 4 and 5, this means $u(V) =
u_{\rm DC} (V) + u_{\rm AC} (V)$, where \beq u_{\rm DC} (V) =
\sum_{{\bf k},m} \tu (k);~ u_{\rm AC} (V) = \frac{1}{V} \sum_{{\bf
k}, m} \sum_{(\bk,m) \neq (\bk',n) \neq (bk,m)} \be_{\bk,m} \cdot
\be_{\bk',n} \sqrt{\tu (k) \tu_(k')} \int e^{i[(\bk - \bk') \cdot
{\bf r} + \phi]} d^3 {\bf r}, \label{uDCAC}\eeq with $\phi =
\phi(\bk,m) - \phi(\bk',n)$, $m,n=$ 1,2 are polarization labels, and
\beq \tu(k) = \frac{\hbar c k}{e^{\beta\hbar k} -1} \label{Uk}\eeq
($\tu$ is the same as $U$ of eq. (\ref{BBRU}) with the phase space
factor removed to enable a full three dimensional mode sum).

The integral in eq. (\ref{uDCAC}) may then be performed.  After
pairing with conjugate terms, we obtain \beq u_{\rm AC} (V) = 16
\sum_{{\bf k}, m} \sum_{(\bk',n) \neq (\bk,m) } ~\be_{\bk,m} \cdot
\be_{\bk',n} \sqrt{\tu (k) \tu_(k')} \frac{\sin(\ka_x L)}{\ka_x L}
\frac{\sin(\ka_y L)}{\ka_y L} \frac{\sin(\ka_z L)}{\ka_z L}
\cos\phi, \label{uACR}\eeq where $\bka = \bk' - \bk$, and it is
understood that for every pair of terms differing from each other by
a mere swap of mode ordering, one member of the pair drops out of
the sum. The variance is \beq (\delta u)^2 = 256 \sum_{{\bf k}, m}
\sum_{(\bk',n) \neq (\bk,m)} \tu (k) \tu_(k')~\<(\be_{\bk,m} \cdot
\be_{\bk',n})^2\> \<\cos^2 \phi\> \left[\frac{\sin(\ka_x L)}{\ka_x
L} \frac{\sin(\ka_y L)}{\ka_y L} \frac{\sin(\ka_z L)}{\ka_z
L}\right]^2. \label{deltau2}\eeq Next, in eq. (\ref{deltau2}) the
sum over $\bk'$ is replaced by $\int d^3 \bk' = \int d^3 \bka = \int
d\ka_x d\ka_y d\ka_z $, and the resulting integration of the three
\texttt{sync}$^{\texttt{2}}$ functions can be done by means of the
reasoning that led to eq. (\ref{deltaIT2}), i.e. for large $L$ only
the range of $\bka$ near $\bka =$ 0 (or $\bk' \approx \bk$)
contributes. The outcome is, \beq (\delta u)^2 = \frac{\pi^2 c^3}{V}
\int_0^\infty d\omega \left[\frac{U(\omega)}{\omega}\right]^2,
\label{deltau22}\eeq where $U(\omega)$ is given by eq. (\ref{BBRU})
and the polarization sum in eq. (\ref{deltau2}) is again carried out
means of eq. (\ref{summn}), i.e. since ${\bf k} \approx {\bf k'}$
the double summation of eq. (\ref{summn}) yields two as the answer,
so that the `helicity factor' of four in $U^2(\omega)$ is reduced to
two. The integral of eq. (\ref{deltau22}) may be recast as
$$(\delta u)^2 = \frac{8\pi k^5 T^5}{h^3 c^3 V} \int_0^\infty
\frac{x^4 dx}{(e^x -1)^2}. $$ Referring back to eqs.
(\ref{deltaeps}) and (\ref{varterms}), and bearing in mind that
$\epsilon= u/V$, we proved that the `non Poisson' (rightmost) term
of eq. (\ref{varterms}), when integrated over all frequencies $x$
according to eq. (\ref{deltaeps}), agrees exactly with the phase
noise variance of the energy density as calculated here.

Further, it should be emphasized that the function $U(\omega)$ in
eq. (\ref{deltau22}) can be the spectral energy density ${\cal U}
(\omega)$ of {\it any} diffuse radiation, and the equation will
still yield the correct phase noise variance.  Moreover, it is also
evident that the Poisson noise (second last) term of eq.
(\ref{varterms}), has the variance of \beq (\delta u)^2 =
\frac{1}{V} \int_0^\infty \hbar\omega {\cal U}(\omega) d\omega,
\label{deltau23}\eeq and the total variance is the sum of the two
contributions, in agreement with the heuristic (maximum Entropy)
treatment of eq. (\ref{deltaU}).

\vspace{2mm}

\noindent {\bf 7. Conclusion: cosmological implications}

Turning finally to the CMB and cosmology, the observed CMB
anisotropy over lengthscales of Mpc or above is far in excess of the
intrinsic 3K BBR noise discussed throughout this paper. It is widely
believed that the origin of the CMB anisotropy is vacuum
fluctuations generated during inflation, and caution must then be
taken when one contemplates those very early epochs.  Thus, as
explained in Lieu \& Kibble (2009), the intrinsic BBR noise at the
end of inflation when the universe was reheated to $kT \sim$
10$^{15}$ eV (GUT) temperatures is by no means negligible; in fact,
on the scale of the GUT horizon the density contrast
$\delta\epsilon/\epsilon$ due solely to this thermal noise (assuming
each island universe is causally connected within its own Hubble
horizon and has therefore the time to reach thermodynamic
equilibrium) is $\approx$ 10$^{-5}$, on par with the observed value
of the same for horizon crossing modes of density perturbation at
relatively much lower redshifts.  This point is particularly
significant since $\delta\epsilon/\epsilon$ on the horizon is, for
the observed Harrison-Zel'dovich form of the power spectrum $P(k)
\sim k$ of superhorizon fluctuations (for the meaning of $P(k)$ see
section 2), both a gauge and epoch independent quantity having a
value set by the initial condition after reheating.

Apart from the above justification, which concerns the {\it
amplitude} of the density seeds, the furthering of our insights on
the BBR noise as developed in this paper also helps us to better
define the role of thermal diffusion.  As case in point, one
important application is to the problem of how $P(k)$ evolves from
an initially spatially homogeneous configuration (the highly ordered
state that ensued from the reheating of a post-inflationary
universe) as a result of the gradual thermalization of superhorizon
regions.  It has been known for some time (Zel'dovich 1965,  Peebles
1974, Zel'dovich and Novikov 1983) that such a process gives rise to
the power spectrum $P(k) \sim k^n$ where $n \geq$ 4.  The inequality
for $n$ is often referred to as the Zel'dovich bound.  However, a
careful examination (Gabrielli et al 2004 and references therein)
reveals that the underlying assumption adopted by these historical
authors is the random walk of classical Poisson-like particles.
Moreover, Gabrielli et al pointed out that if the diffusion
mechanism at work in the very early universe involves also the
`fuzzy' wave-like properties of its constituent `particles' -
properties that `blurs' the sharpness of the cutoff of interactions
beyond the horizon separation distance because the `particles'
cannot be localized as readily - the Zel'dovich bound might be
broken, opening the range of allowed $n$ to even include the
Harrison-Zel'dovich value of $n \approx$ 1. Thus, in this context
the present paper clearly plays a role in bolstering the argument of
Gabrielli et al, viz. a radiation dominated ensemble {\it does}
exhibit density fluctuations that originate from the phase
interference of waves, hence it is by no means a system of
classically diffusing particles. As explained in Lieu \& Kibble
(2009), under the circumstance the attractive prospect of accounting
for {\it both} the amplitude and index of primordial density
perturbations without invoking fluctuations in the vacuum energy
then avails itself.

\noindent {\bf 8. References}

\noindent Bennett, C.L. et al, 2003, ApJS, 148, 97.

\noindent  Gabrielli, A., Joyce, M., Marcos, B., \& Viot, P., 2004,
Europhys. Lett., 66, 1.


\noindent Hanbury Brown, R., and Twiss, R.Q., 1957, Proc. Roy. Soc.
A, 242, 300.


\noindent Hinshaw, G. et al 2007, ApJS, 170, 288.

\noindent Jaynes, E.T. 1957, Physical Review, 106, 620.

\noindent Lieu, R. \& Kibble, T.W.B. 2009, ApJL submitted
(arXiv:0904.4840).

\noindent Mandel, L., 1958, Proc. Phys. Soc. 72 1037.

\noindent Meszaros, P., 1974, A \& A, 37, 225.

\noindent Peacock, J.A., 1999, Cosmological Physics, Cambridge
University Press.

\noindent Peebles, P.J.E., 1974, A \& A, 32, 391.

\noindent Peebles, P.J.E., 1993, Principles of Physical Cosmology,
Princeton University Press.

\noindent Purcell, E.M., 1956, Nature, 178, 1449.

\noindent Spergel, D.N. et al 2007, ApJS, 170, 377.

\noindent Trebino, R., et al, 1997, Rev. Sci. Instr., 68, 3277.

\noindent Zel'dovich, Ya., 1965, Adv. Astron. Ap. 3, 241.

\noindent Zel'dovich, Ya., \& Novikov, I., 1983, Relativistic
Astrophysics, Vol. 2, Univ. Chicago Press.

\end{document}